\documentclass[superscriptaddress,twocolumn,pre]{revtex4}
\usepackage{amssymb}
\usepackage{natbib}
\usepackage{amsmath}
\usepackage{amsfonts}
\usepackage{graphicx}
\usepackage{mathrsfs}
\usepackage{dcolumn} 
\usepackage{bm}
\usepackage{color}
\usepackage{cancel}
\usepackage{mathbbol}
\usepackage{epsfig}
\usepackage{units}
\usepackage{esint}
\usepackage{soul} 
\usepackage{url}

\definecolor{rot}{rgb}{0.75,0.05,0.25}
\definecolor{hellgrau}{gray}{0.5}
\definecolor{blau}{rgb}{0,0,0.7}

\def\Tr{\mbox{Tr}}

\begin{document}

\title[]{Nonadiabatic single-qubit quantum Otto engine}

\author{Andrea Solfanelli}
\address{Department of Physics and Astronomy, University of Florence, Via Sansone 1, I-50019, Sesto Fiorentino (FI), Italy.}
\address{INFN Sezione di Firenze, via G. Sansone 1, I-50019, Sesto Fiorentino (FI), Italy.}
\author{Marco Falsetti}
\address{Department of Physics and Astronomy, University of Florence, Via Sansone 1, I-50019, Sesto Fiorentino (FI), Italy.}
\author{Michele Campisi}
\address{Department of Physics and Astronomy, University of Florence, Via Sansone 1, I-50019, Sesto Fiorentino (FI), Italy.}
\address{INFN Sezione di Firenze, via G. Sansone 1, I-50019, Sesto Fiorentino (FI), Italy.}

\begin{abstract}
According to Clausius formulation of the second law of thermodynamics, for any thermal machine withdrawing heats $Q_{1,2}$ from  two heat reservoirs at temperatures $T_{1,2}$, it holds $Q_1/T_1+Q_2/T_2 \leq 0$. Combined with the observation that the quantity $Q_1+Q_2$ is the work $W$ done by the system, that inequality tells that only 4 possible operation modes are possible for the thermal machine, namely heat engine [E], refrigerator [R], thermal accelerator [A] and heater [H]. We illustrate their emergence in the finite time operation of a quantum Otto engine realised with a single qubit. We first focus on the ideal case when isothermal and thermally-insulated strokes are well separated, and give general results as well as results pertaining to the specific  finite-time Landau-Zener dynamics. We then present realistic results pertaining to the solid-state  experimental implementation proposed by Karimi and Pekola [Phys. Rev. B \textbf{94} (2016) 184503]. That device is non-adiabatic both in the quantum mechanical sense and in the  thermodynamical sense. Oscillations in the power extracted from the baths due to coherent LZ tunnelling at too low temperatures are observed that might hinder the robustness of the operation of the device against experimental noise on the control parameters.
\end{abstract}

\maketitle

\section{Introduction}
One of the cornerstones of thermodynamics is Clausius inequality:
\begin{align}
\sum_i \frac{Q_i}{T_i} \leq 0 \, ,
\label{eq:clausius}
\end{align}
where $Q_i$ are the energies that a central system undergoing a cyclic transformation withdraws from a set of surrounding heat baths at temperatures $T_i$. In the continuum limit it gives the celebrated expression $\oint \delta Q/T \leq 0$. Noting that the equal sign holds when the cycle is reversible, led Clausius to naturally introduce the state function $S$ such that $dS=\delta Q_\text{rev}/T$, with $Q_\text{rev}$ being the heat exchanged during a reversible transformation, and the property that $S(B)-S(A)\geq \int_A^B \delta Q/T$. $S$ is the entropy \cite{Fermi56Book}.

\begin{figure}
		\begin{center}
		\includegraphics[width=\linewidth]{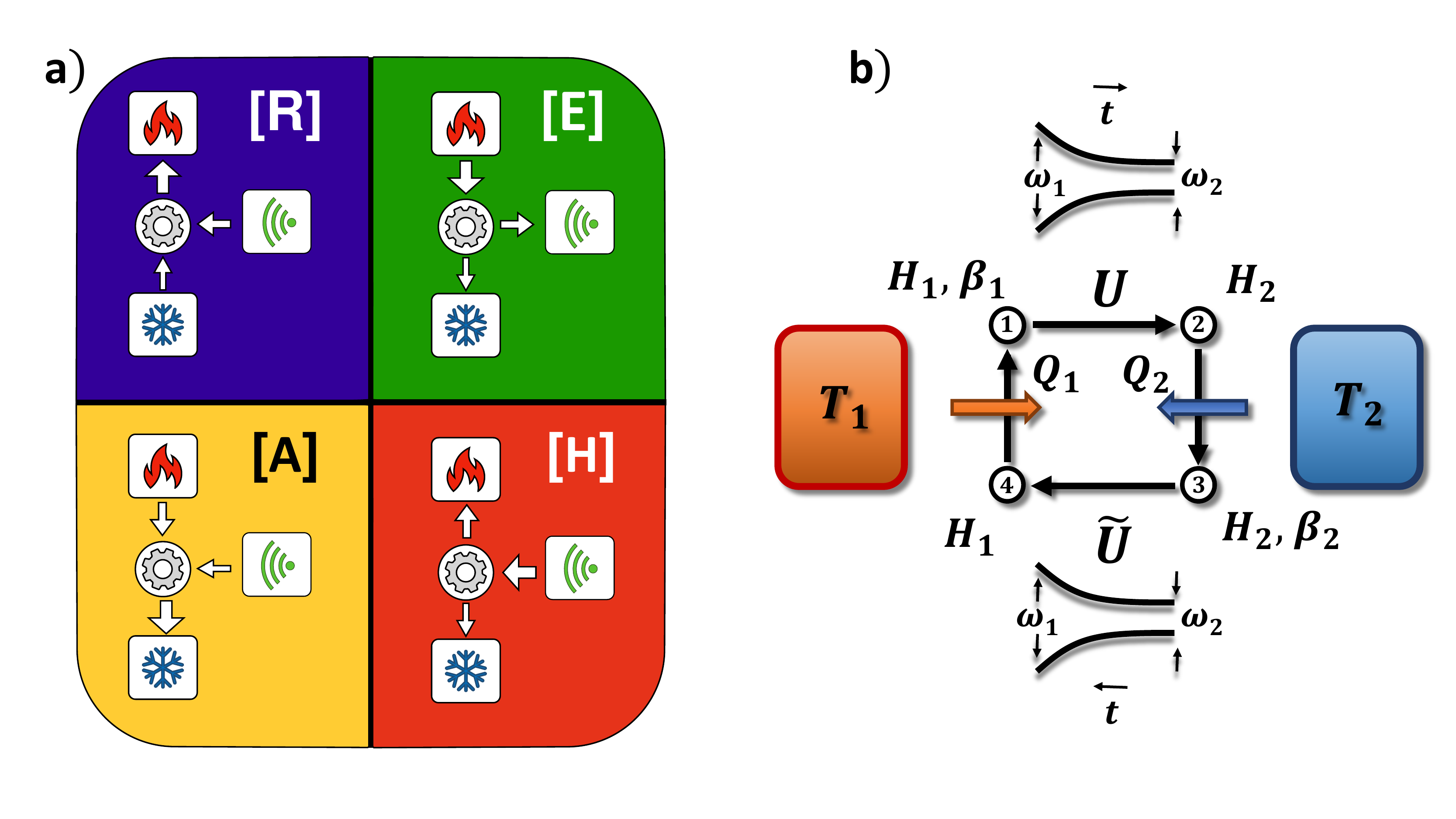}
		\caption{Panel a) The four possible operation modes for a device working with two reservoirs. Panel b) Sketch of a single-qubit based quantum Otto engine cycle.}
		\label{fig:1}
		\end{center}
\end{figure}

For the simplest case of a central system interacting with two baths only and a work source, the above Eq. (\ref{eq:clausius}), combined with the first law of thermodynamics, gives the following conditions
\begin{align}
\beta_1Q_1+ \beta_2 Q_2 &\leq 0 \label{eq:ClausiusFor2}\\
Q_1+Q_2 &= W \, , \label{eq:NRGconservation}
\end{align}
where $W$ is the work delivered to the work source and $\beta_i$ are the baths inverse thermal energies $\beta_i=(k T_i)^{-1}$,  and $k$ is Boltzmann's constant.
Looking at the system as a thermal machine, depending on the signs of $Q_1,Q_2,W$ it may realise various operation modes. 
Basic mathematics show that the above constraints are simultaneously compatible only with 4 (out of a total of $8=2^3$) operation modes, see Appendix \ref{app:A}. Setting, without loss of generality, the convention $\beta_1< \beta_2$, these are 
\begin{align}
\text{[R]: }&  Q_1\leq 0 \quad Q_2\geq0 \quad W\leq0 \label{eq:R}\\
\text{[E]: }& Q_1\geq0 \quad  Q_2\leq 0 \quad W \geq 0 \label{eq:E}\\
\text{[A]: }&  Q_1\geq0 \quad Q_2\leq0 \quad W \leq0 \label{eq:A}\\
\text{[H]: }&  Q_1\leq0 \quad Q_2\leq0 \quad W \leq0 \label{eq:H}\, ,
\end{align}
where [E] denotes energy extraction (heat engine), [R] denotes refrigerator, [A] denotes thermal accelerator, and [H] denotes heater \cite{Buffoni19PRL122}. They are illustrated in Fig. \ref{fig:1}, panel a).

In this work we illustrate how the four operations emerge in a quantum Otto engine operating in finite time.

\section{Single-qubit quantum Otto engine}
We consider an engine consisting of a single qubit undergoing a four stroke cycle. See Fig. \ref{fig:1} panel b).
We assume the qubit is initially at thermal equilibrium at temperature $T_1$.
In the first stroke the qubit undergoes an evolution where its resonant frequency changes from a value $\omega_1$ to a value $\omega_2$. In the second stroke the qubit at fixed resonant frequency $\omega_2$ interacts with the thermal bath $2$ so as to reach temperature $T_2$. In the third stroke the qubit undergoes a reversed evolution where its resonant frequency changes from the value $\omega_2$ to the value $\omega_1$. In the fourth stroke the qubit at fixed resonant frequency $\omega_1$ interacts with the thermal bath $1$ so as to reach temperature $T_1$, thus closing the cycle.

The first and third strokes are adiabatic in a thermodynamic sense (namely they occur in thermal isolation), but are not necessarily adiabatic in the quantum-mechanical sense, namely, during the evolution quantum transitions may occur.

The unitary dynamics $U$ occurring in the first stroke are generated by a generic time-dependent spin-${1}/{2}$ Hamiltonian $H(t)$,  according to the rules of quantum mechanics:
\begin{align}
H(t) &= x(t) \sigma_x + y(t)\sigma_y +z(t)\sigma_z,  \quad t \in [t_1,t_2] \, ,\\
U &= \text{T}\exp\left[-(i/\hbar) \int_{t_1}^{t_2} H(s)ds \right]\, ,
\label{eq:UF}
\end{align}
where $ \text{T}\exp$ denotes the time-ordered exponential, $t_1$ and $t_2$ are initial and final times of the stroke, $\sigma_{x,y,z}$ denote Pauli operators, and $x(t),y(t),z(t)$ are generic time dependent real coefficients.
The qubit level spacings at the beginning and end of the first stroke are give by the expressions:
\begin{align}
\hbar \omega_1&=2\sqrt{x^2(t_1)+y^2(t_1)+z^2(t_1)},  \\
\hbar \omega_2&=2\sqrt{x^2(t_2)+y^2(t_2)+z^2(t_2)}
\end{align}

The unitary dynamics $\widetilde{U}$ occurring  in the third stroke are generated by the time-dependent Hamiltonian realising the ``backward'' protocol
\begin{align}
\widetilde{H}(t) &= H(t_2+t_1-t), \quad t \in [t_1,t_2] \\
\widetilde{U} &= \text{T}\exp\left[-(i/\hbar) \int_{t_1}^{t_2} \widetilde{H}(s)ds \right] \, .
\label{eq:UB}
\end{align}
We assume that at each time $t\in[t_1,t_2]$ the Hamiltonian $H(t)$ is invariant under the action of some anti-unitary operator $K$, $H(t)=K H(t)^\dagger K^\dagger$ \footnote{Usually one assumes $K$ to be the time-reversal operator \cite{Campisi11RMP83}, but that is not necessary for our purposes.}. Under that assumption it is  \cite{Schmidtke18PRE98}:
\begin{align}
\widetilde{U} = K U^\dagger K^\dagger \, .
\end{align}
For simplicity we introduce the notation:
\begin{align}
H_1 &= H(t_1) = \widetilde{H}(t_2)\\
H_2 &= H(t_2) = \widetilde{H}(t_1)\, .
\end{align}

Let $\rho_i$ and $E_i$ denote the state of the qubit at the beginning of stroke $i$ and its according energy expectation value. By the thermalisation assumption and Eqs. (\ref{eq:UF},\ref{eq:UB}),  it  is:
\begin{align}
\left.\begin{array}{ll}
\rho_1 = e^{-\beta_1 H_1}/Z_1, \qquad & E_1=\Tr \rho_1H_1 \\
\rho_2 = U \rho_1 U^\dagger, \qquad & E_2=\Tr \rho_2H_2 \\
\rho_3 = e^{-\beta_2 H_2}/Z_2, \qquad & E_3=\Tr \rho_3H_2 \\
\rho_4 = \widetilde{U} \rho_3 \widetilde{U}^\dagger, \qquad & E_4=\Tr \rho_4H_1 \, ,
\end{array}\right.
\end{align}
where $Z_i=\Tr e^{-\beta_i H_i}$ is the canonical partition function.

The thermodynamics of the engine is fully characterised by the heats $Q_{1},Q_{2}$ withdrawn from the baths  $1,2$ during the thermalisation steps which, under the assumption of weak qubit bath coupling, equal the energies gained by the qubit during those strokes namely:
\begin{align} 
Q_{2} &= E_3-E_2 \\
Q_{1} &= E_1-E_4\, .
\end{align}
Since $E_1$ and $E_3$ are thermal expectations they can be readily expressed as:
\begin{align}
E_1 &= -\frac{\hbar \omega_1}{2} \tanh\left (\frac{\beta_1 \hbar \omega_1}{2} \right)\\
E_3 &= -\frac{\hbar \omega_2}{2} \tanh\left (\frac{\beta_2 \hbar \omega_2}{2} \right)\, .
\end{align}
For $E_2$ we obtain:
\begin{align}
E_2 =
\frac{\sum_{i,j} \varepsilon_i^{(2)}  e^{- \beta_1 \varepsilon_j^{(1)}} | \langle \psi_i^{(2)} |U |\psi_j ^{(1)} \rangle|^2   }{2\cosh(\beta_1\hbar \omega_1/2)}\, ,
\end{align}
where $|\psi_i^{(r)}\rangle, \varepsilon_i^{(r)}$, $i,r=1,2$ are the eigenvectors and corresponding eigenvalues of $H_r$: $H_r|\psi_
i^{(r)}\rangle= \varepsilon_i^{(r)} |\psi_i^{(r)}\rangle$. Chosing the label $i=1,2$ for the ground  and excited states respectively, it is $\varepsilon_1^{(r)}=-\hbar \omega_r/2, \varepsilon_2^{(r)}=+\hbar \omega_r/2$.

Note that the $2\times 2$ square matrix $P_{ij}= | \langle \psi_i^{(2)} |U |\psi_j ^{(1)} \rangle|^2 $ is doubly stochastic \cite{Marshall11book}, namely $0\leq P_{ij}\leq 1$, $\sum_i P_{ij}=\sum_j P_{ij}=1$. This immediately implies that one of its elements is sufficient to determine all of them, and that the matrix is symmetric:
if $P_{11}\doteq P$, then $P_{12}=P_{21}=1-P$, and $P_{22}=P$. Then $E_2$ reads
\begin{align}
E_2 =
\frac{\hbar \omega_2}{2} \tanh\left (\frac{\beta_1 \hbar \omega_1}{2} \right)(1-2P) \, ,
\end{align}
where $P$ contains all relevant information pertaining to the degree of adiabaticity of the sweep. In the adiabatic limit where there occur no transitions among the instantaneous energy eigenstates, we have $P\to 1$ and $E_2 \to E_1 \omega_2/\omega_1$ in accordance with the energy spectrum getting dilated/contracted by a factor $\omega_2/\omega_1$.

Similarly, for the calculation of $E_4$ we obtain
\begin{align}
E_4 =
\frac{\sum_{i,j} \varepsilon_i^{(1)}  e^{- \beta_2 \varepsilon_j^{(2)}} | \langle \psi_i^{(1)} |\widetilde{U} |\psi_j ^{(2)} \rangle|^2   }{2\cosh(\beta_2\hbar \omega_2/2)}\, .
\end{align}
Using Eq. (\ref{eq:UB}) and the property $\langle u | K A K^\dagger |w\rangle = \langle u | A |w\rangle^*$ (with $A$ a linear operator, and $K$ an anti-linear operator) \cite{Messiah62Book} we get
\begin{align}
 &| \langle \psi_i^{(1)} |\widetilde{U} |\psi_j ^{(2)} \rangle|^2 
 =  | \langle \psi_i^{(1)} | K U^\dagger K^\dagger |\psi_j ^{(2)} \rangle|^2 =  \nonumber \\
& | \langle \psi_i^{(1)} | U^\dagger |\psi_j ^{(2)} \rangle^*|^2 
 =  | \langle \psi_j^{(2)} | U |\psi_i ^{(1)} \rangle|^2=P_{ji}=P_{ij}\, ,
\end{align}
hence
\begin{align}
E_4 =
\frac{\hbar \omega_1}{2} \tanh\left (\frac{\beta_2 \hbar \omega_2}{2} \right)(1-2P)\, .
\end{align}
Summing up:
\begin{align} 
Q_{1} &= -\frac{\hbar \omega_1}{2} \left [\tanh\left (\frac{\beta_1 \hbar \omega_1}{2} \right)+\tanh\left (\frac{\beta_2 \hbar \omega_2}{2} \right)(1-2P)\right]
\label{eq:Q1}\\
Q_{2} &= -\frac{\hbar \omega_2}{2} \left [\tanh\left (\frac{\beta_2 \hbar \omega_2}{2} \right)
+\tanh\left (\frac{\beta_1 \hbar \omega_1}{2} \right)(1-2P)\right]
\label{eq:Q2}\\
W &= - \frac{\hbar}{2} \tanh\left (\frac{\beta_1 \hbar \omega_1}{2} \right)[\omega_1+\omega_2(1-2P)] \nonumber\\
& -\frac{\hbar}{2} \tanh\left (\frac{\beta_2 \hbar \omega_2}{2} \right)[\omega_1(1-2P)+ \omega_2]\, .
\label{eq:W}
\end{align}
The transition probability $P$ contains information pertaining to the qubit dynamics during the unitary strokes, as such it is a functional of $\{x(t),y(t),z(t)\}$, which we leave unspecified for now.
It is trivial to note that all energy exchanges are increasing functions of $P$ and are maximal 
in the adiabatic limit $P=1$.
Departure from that limit means smaller exchanges. As we shall see below it also means smaller thermodynamic efficiencies.

It is not difficult to see that, with $Q_{1,2}$ as in Eq. (\ref{eq:Q1},\ref{eq:Q2}) it is $\beta_1 Q_1 + \beta_2 Q_2 \leq 0$ in accordance with Eq. (\ref{eq:ClausiusFor2}), see Appendix \ref{app:B}.

Since the above expressions are linear in $P$, it is easy to find the values of $P$, call them $P_{Q_1},P_{Q_2},P_W$ for which $Q_1,Q_2$ and $W$, respectively, become null, and therefore mark their sign reversal:
\begin{align} 
P_{Q_{1}} &=\frac{1}{2}\left [ 1+\frac{\tanh(\beta_1\hbar \omega_1/2) }{\tanh(\beta_2\hbar \omega_2/2) }\right]  \label{eq:PQ1} \\
P_{Q_{2}} &= \frac{1}{2}\left [ 1+\frac{\tanh(\beta_2\hbar \omega_2/2)}{\tanh(\beta_1\hbar \omega_1/2)}\right] \\
P_{W}&= \frac{1}{2}\left [ 1+\frac{\hbar \omega_1\tanh(\beta_1\hbar \omega_1/2)+ \hbar \omega_2\tanh(\beta_2\hbar \omega_2/2)}
{\hbar \omega_2 \tanh(\beta_1\hbar \omega_1/2) + \hbar \omega_1 \tanh(\beta_2\hbar \omega_2/2)}\right]\, .
\end{align}

\begin{figure}
		\begin{center}
		\includegraphics[width=\linewidth]{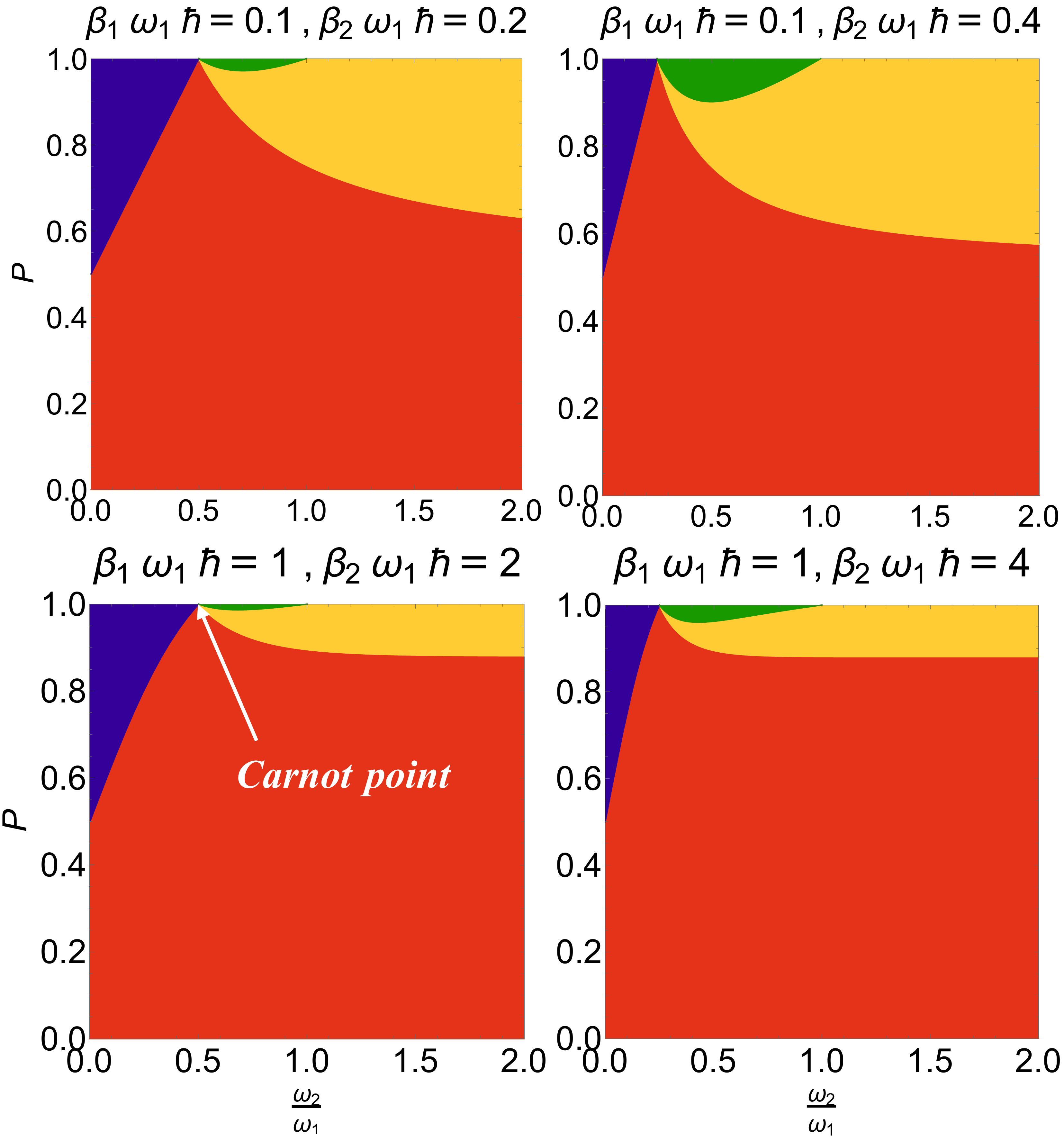}
		\caption{Operation regions in the $(\omega_2/\omega_1,P)$ plane. Blue=[R]. Red=[H]. Green = [E]. Yellow = [A], in accordance with the color convention in Fig \ref{fig:1} a). 	
		Moving downward the ratio $\beta_1/\beta_2$ is fixed, and $\beta_2$ increases. Moving to the right $\beta_1$ is fixed, while $\beta_1/\beta_2$ decreases. The Carnot point $(\omega_2/\omega_1=\beta_1/\beta_2,P=1)$ where all operations coincide is indicated in panel c) only.}
		\label{fig:2}
		\end{center}
\end{figure}
\begin{figure*}
		\begin{center}
		\includegraphics[width=\linewidth]{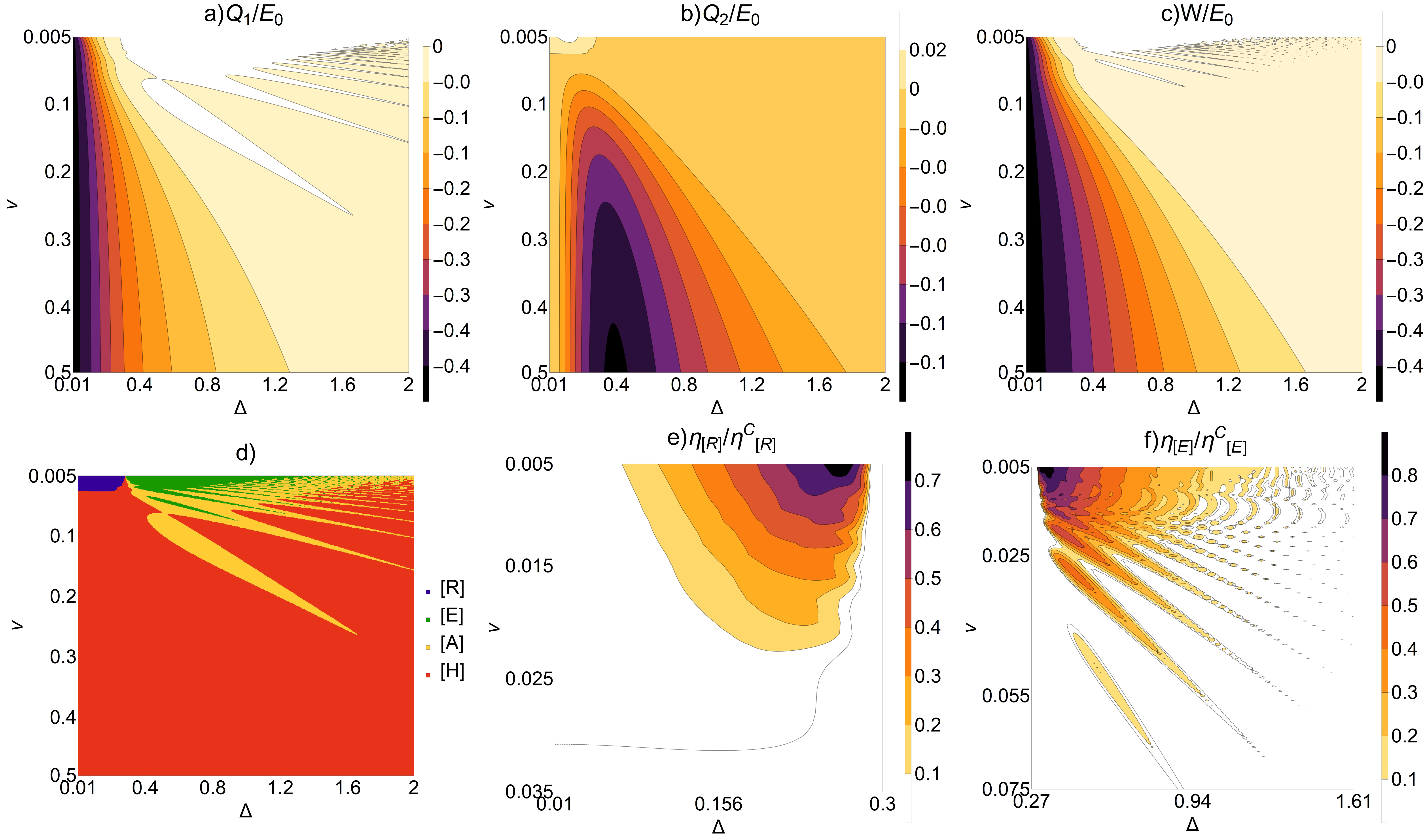}
		\caption{Thermodynamics of the Landau-Zener-St\"uckelberg-Majorana quantum Otto engine as function of the nondimensional parameters $v,\Delta$ at fixed $v\tau=1/2$. Panel a): Heat withdrawn from resistor 1 in one cycle.  Panel b) Heat withdrawn from resistor 2 in one cycle. Panel c) Work output. Panel d) Operation regions: Blue=[R]; Red=[H]; Green = [E]; Yellow = [A]. Panel e): Rescaled refrigeration efficiency $\eta_{[R]}/\eta^C_{[R]}$. Panel f):  Rescaled heat engine efficiency $\eta_{[E]}/\eta^C_{[E]}$. The temperature ratio is $T_1/T_2=\beta_2/\beta_1=2$, while $\beta_1E_0 =10/3 $. The Carnot point is accordingly at $v=0, \Delta = 1/\sqrt{12}\simeq0.29$, and the Carnot efficiencies are $\eta_{[E]}^{C} = 1/2$ and $\eta_{[R]}^{C} =1 $. }
		\label{fig:3}
		\end{center}
\end{figure*}
Figure \ref{fig:2} shows the curves $P_{Q_1},P_{Q_2},P_W$, as a function of $\omega_2/\omega_1$, for various fixed values of $\beta_1\hbar \omega_1,\beta_2 \hbar \omega_1$. Crossing one curve means reversing the sign of the according quantity, therefore the curves draw the boundaries of the regions of distinct operation modes. As expected there are only four regions,  corresponding each to one of the four allowed operation modes described above, which we have filled with different colours.

A few observations are in order. First, the regions, as plotted in the $(\omega_2/\omega_1, P)$ plane are connected. We note that for $P<1/2$ only the [H] operation is possible: from Eqs. (\ref{eq:Q1}, \ref{eq:Q2}) one can immediately see that for $P<1/2$ it is $Q_1\leq 0, Q_2 \leq 0$. As $P$ gets larger and larger, above the value $1/2$ the [H]-region shrinks until it reduces to a single point, which is in fact the Carnot point (characterised by $\omega_2/\omega_1 = \beta_1/\beta_2$ and $P=1$) where all operations coincide, and actually nothing happens (i.e., all exchanged energies are null). For $P=1$, corresponding to the  quasi-static (i.e., adiabatic in the quantum-mechanical sense) limit, only [R], [E] and [A] are possible, with [R] occurring for $\omega_2/\omega_1 \leq \beta_1/\beta_2$, $[E]$ occurring for $\beta_1/\beta_2 \leq \omega_2/\omega_1 \leq 1 $, and [A] occurring for $\omega_2/\omega_1 \geq 1$. As $P$ decreases from the value $1$, the according intervals become smaller at the expense of an enlarging [H] interval. 

Note that $P_{Q_2}\to 1/2$ as $\hbar \omega_2 \to 0$, which gives the [R] region a triangular-like shape, with one side of fixed length and the other side  with a size that gets smaller and smaller as the ratio $\beta_1/\beta_2$ decreases. That reflects the fact that, at fixed hot temperature $T_1$, extracting heat from the cold bath becomes more and more difficult as $T_2$ decreases. Note also that contextually, the [E] region would expand, reflecting the fact that it is easier to deliver positive work, when there is a larger thermal gradient. The region [H] is the biggest in the $(\omega_2/\omega_1, P)$ plane, in accordance with the intuitive idea that dumping heat in both baths is the generally the easiest thing to accomplish.

As anticipated, transitions are responsible for drops in the thermodynamic efficiency.
To see that, consider the [E] regime, where $Q_1\geq 0,Q_2\leq 0$. The thermodynamic efficiency $\eta_{[E]}=W/Q_1=1+Q_2/Q_1$ reads:
\begin{align} 
\eta_{[E]} = 1+ 
\frac{\omega_2}{\omega_1} \frac
{\tanh(\beta_2\hbar \omega_2/2)+\tanh(\beta_1\hbar \omega_1/2)(1-2P)}
{\tanh(\beta_1\hbar \omega_1/2)+\tanh(\beta_2\hbar \omega_2/2)(1-2P)}
\end{align}
Note that both numerator and denumerator in the equation above decrease with increasing $P$, so while the numerator (i.e., $-2 Q_2/\hbar$, which is positive) becomes less positive, the denumerator (i.e., $-2 Q_1/\hbar$ which is negative) becomes more negative. Accordingly, the absolute value of the ratio (which is a negative number) decreases with increasing $P$, implying that $\eta_{[E]}$ is an increasing function of $P$. Similarly one can see that  the coefficient of performance in the [R]-operation decreases with $P$. Accordingly best thermodynamic performances are achieved in the quasi-static limit, where the quasi-static Otto efficiencies  \cite{Quan07PRE76,DelCampo14SCIREP4} occur:
$
\eta_{[E]}^{qs} = 1- \omega_2/\omega_1
$
and $\eta_{[R]}^{qs} = 1/(\omega_1/\omega_2-1)$, which in turn increase and tend to the Carnot efficiencies $
\eta_{[E]}^{C} = 1- \beta_1/\beta_2
$, $\eta_{[R]}^{C} = 1/(\beta_2/\beta_2-1)$ as one gets close to the Carnot point. However, as pointed out above the absolute value of the exchanged energies go to zero at the Carnot point.

\section{Landau-Zener-St\"uckelberg- Majorana dynamics}
\label{sec:LZSM}
\begin{figure*}
		\begin{center}
		\includegraphics[width=\linewidth]{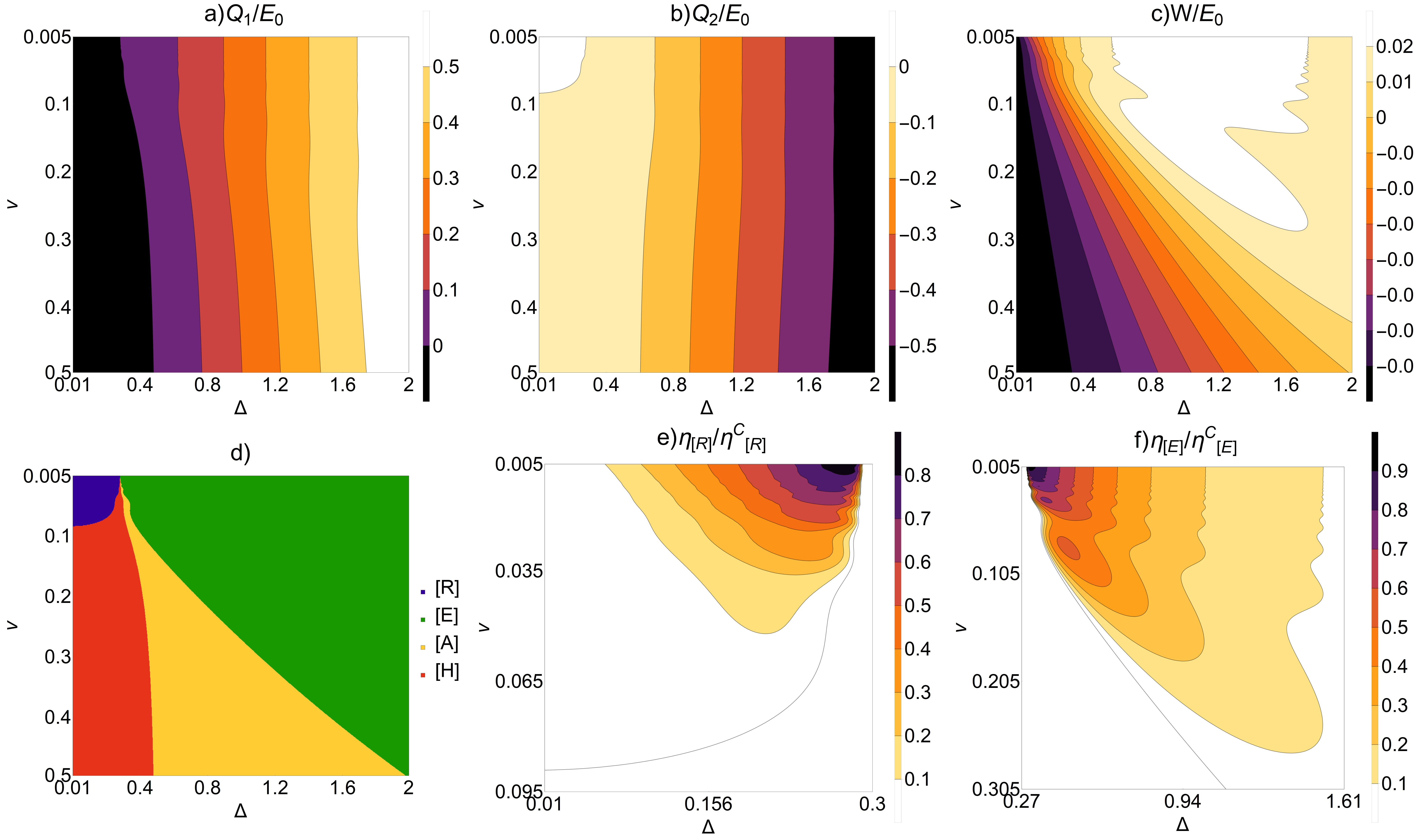}
				\caption{Same as Fig. \ref{fig:3} but for $\beta_1 E_0=1/3$.}
		\label{fig:4}
		\end{center}
\end{figure*}

We now focus on the specific case of the Landau-Zener-St\"uckelberg-Majorana \cite{Landau32PZS2,Zener32PRSA137,Stueckelberg32HPA5,Majorana32NC9} evolution, 
\begin{align}
x(t)= \delta,\quad y(t)=0, \quad z(t)= u t
\end{align}
Note that, if adopting the $\{\sigma_z\}$ representation where the Pauli matrices $\sigma_{x},\sigma_{z}$ are real, the according Hamiltonian is invariant under the anti-unitary complex conjugation $K_{\{\sigma_z\}}$ relative to that representation \cite{Messiah62Book}, so we are within the assumptions stated above.

The transition probability $P$ can be expressed in this case in terms of the problem parameters $(\delta, u, t_1, t_2)$ by means of parabolic cylinder functions, as described in \cite{Vitanov96PRA53}. The unitary evolution in the $\{\sigma_z\}$ representation reads:
\begin{align}
U_{11}(t_2,t_1)&=U^*_{22}(t_2,t_1)= \frac{\Gamma(1-i\frac{\delta^2}{2u})}{\sqrt{2\pi}}\times \nonumber\\
&\left[ D_{i\frac{\delta^2}{2u}}(t_2\sqrt{2u}e^{-i\pi/4}) D_{i\frac{\delta^2}{2u}-1}(t_1\sqrt{2u}e^{i3\pi/4})+ \right.\nonumber\\
&\left. D_{i\frac{\delta^2}{2u}}(t_2\sqrt{2u}e^{i3\pi/4}) D_{i\frac{\delta^2}{2u}-1}(t_1\sqrt{2u}e^{-i\pi/4}) \right] \\
U_{12}(t_2,t_1)&=-U^*_{21}(t_2,t_1)=   \frac{\Gamma(1-i\frac{\delta^2}{2u})e^{i\pi/4}}{\delta\sqrt{\pi/u}}\times \nonumber\\
& \left[-D_{i\frac{\delta^2}{2u}}(t_2\sqrt{2u}e^{-i\pi/4}) D_{i\frac{\delta^2}{2u}}(t_1\sqrt{2u}e^{i3\pi/4})+\right. \nonumber\\
& \left. D_{i\frac{\delta^2}{2u}}(t_2\sqrt{2u}e^{i3\pi/4}) D_{i\frac{\delta^2}{2u}}(t_1\sqrt{2u}e^{-i\pi/4}) \right]
\end{align}
where $U_{ij}=\langle \phi_i|U| \phi_j\rangle$ and $|\phi_{1}\rangle, |\phi_{2}\rangle$ denote the spin-down and spin-up eigenvectors of $\sigma_z$, respectively;
$D_\nu(z)$ denotes the parabolic cylinder D-function and $\Gamma(z)$  is the Gamma Function \cite{Gradshteyn07Book}. 
Denoting with $\mathbf{U}^{ad}(t_2,t_1)$ the time evolution matrix expressed in the time varying instantaneous Hamiltonian eigenbasis, it is \cite{Vitanov96PRA53}:
\begin{align}
\mathbf{U}^{ad}(t_2,t_1)=
\mathbf{R}^T(t_2)\cdot \mathbf{U}(t_2,t_1) \cdot \mathbf{R}(t_1) ,
\end{align}
where $\mathbf{U}(t_2,t_1)$ is the time evolution matrix expressed in the $\{\sigma_z\}$ eigenbasis, $\mathbf{R}(t_r)$, $r=1,2$ is the rotation that changes the basis from adiabatic (eigenstates of $H_r$) to diabatic (eigenstates of $\sigma_z$):
\begin{align}
\mathbf{R}(t)=
\left(\begin{array}{cc} \cos\vartheta(t) &  \sin\vartheta(t)\\ -\sin\vartheta(t) &  \cos\vartheta(t) \end{array}\right), \vartheta(t)=\frac{1}{2}\arctan\frac{\delta}{u t} .
\end{align}
The probability $P$ then reads $P=|\mathbf{U}_{11}^{ad}(t_2,t_1)|^2$. 

Introducing a qubit energy scale $E_0$, we define the nondimensional level spacing $\Delta= \delta/E_0$, nondimensional sweep rate, $v=u\hbar/E_0^2$ and nondimensional time $s=E_0t/\hbar$.

In the following we focus on a sweep between nondimensional time $s_1=-\tau$ (with $\tau>0$) and $s_2=0$, corresponding to $\hbar \omega_1=2E_0 \sqrt{\Delta^2+(v \tau)^2}$, $\hbar \omega_2=2\Delta E_0$. Note that with this choice it is $\omega_2\leq \omega_1$, hence we are exploring the region where all four operations may occur (for $\omega_2\geq \omega_1$ only [H] and [A] may occur).
Figures \ref{fig:3},\ref{fig:4} show, for fixed temperature ratio $\beta_2/\beta_1$ and different $\beta_1$, the plots of $Q_1/E_0,Q_2/E_0,W/E_0$, the regions of operations and the rescaled thermodynamic efficiencies $\eta_{[R]}/\eta^C_{[R]}$ and $\eta_{[E]}/\eta^C_{[E]}$,
 as a function of $v,\Delta$, for fixed $\alpha=v\tau$. The limit $v\to0$ ($\tau\to \infty$) corresponds to the adiabatic limit. Note how the low temperature plots (Fig. \ref{fig:3}) present oscillations in $\Delta$ and $v$, which result in a breakdown of the connectedness of the operation regions. These are a consequence of the well-known oscillations that characterise LZSM transitions in finite time \cite{Vitanov96PRA53,Vitanov99PRA59}. Note that they do not appear in the higher temperature plots (Fig. \ref{fig:4}). Note how, as $\Delta$ increases the intervals of $v$ where [A] occurs shrink, and in practice only [H] occurs in the $\Delta \to \infty$ limit. This is well visible in low temperature plot (Fig. \ref{fig:3} panel d), and would be visibile at higher temperatures (Fig. \ref{fig:4} panel d), if one would enlarge the $\Delta$-axis end-scale accordingly. This behaviour can be understood by looking at Eq. (\ref{eq:PQ1}). When $\Delta \rightarrow \infty$ both $\omega_1$ and $\omega_2$ go to infinity, hence the reversal point $P_{Q_1}$ goes to $1$, meaning that only at $P=1$ (that is in the slow limit $v\rightarrow 0$), [A] can occur. Physically the reason is that when $\Delta$ is very large, almost all qubit population is in the ground state at the beginning of an adiabatic stroke, hence the qubit jumping-up during the adiabatic stroke has an overwhelmingly larger probability than jumping-down. Accordingly the probability of releasing energy to the bath in the subsequent thermalisation stroke is overwhelmingly larger than the probability of withdrawing it.

\section{The Quantum Otto Engine of Karimi and Pekola}
\label{sec:KP}

\begin{figure}
		\begin{center}
		\includegraphics[width=\linewidth]{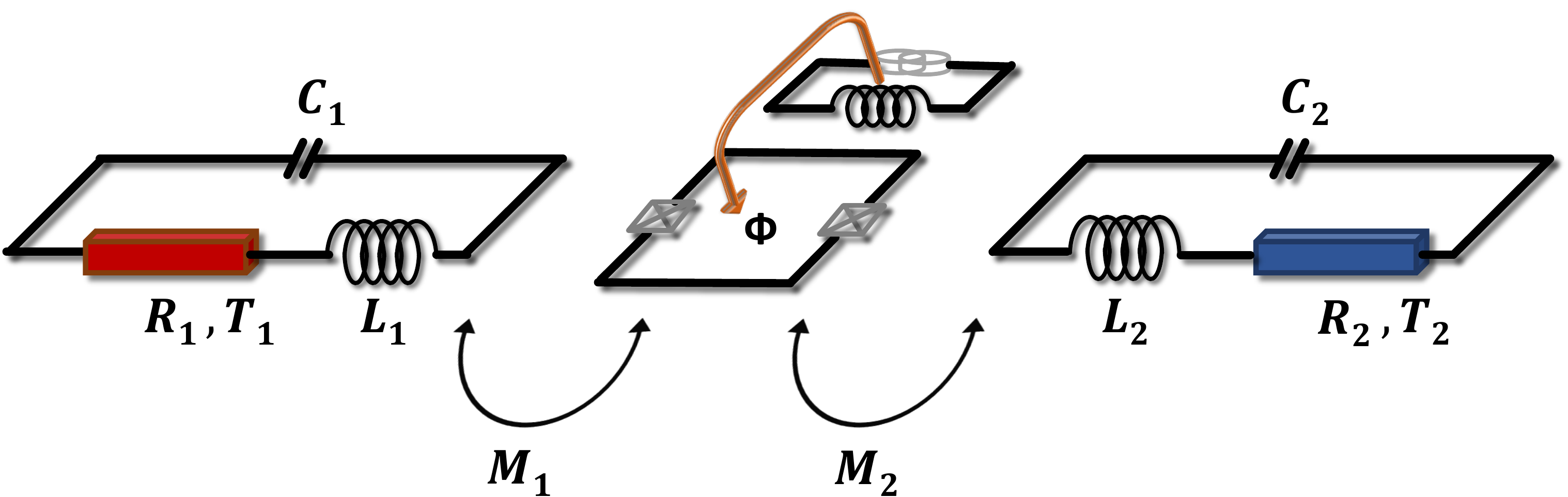}
		\caption{Circuit scheme of the Quantum Otto Engine of Karimi and Pekola \cite{Karimi16PRB94}.}
		\label{fig:5}
		\end{center}
\end{figure}

We consider the solid state implementation of a quantum Otto engine presented by Karimi and Pekola \cite{Karimi16PRB94} whereby a superconducting qubit is coupled to two resistors $R_{j}, j=1,2$ each kept at temperature $T_{j}$ (inverse thermal energy $\beta_j=1/(kT_j)$) and embedded each into an RLC circuit with  resonant frequency $\omega_{LC,j}=1/\sqrt{L_jC_j}$ with $L_j,C_j$ the $j$-th circuit inductance and capacitance, respectively. The coupling between the qubit and resistor $R_j$ is achieved by tuning the qubit level spacing to $\omega_{LC,j}$. Control over the qubit level spacing is provided through the control over the magnetic flux, $\Phi$, that threads the qubit, and is generated by a nearby inductor. The Otto engine is then realised by bringing the qubit in tune with the two LRC circuits, alternatively, see Fig. \ref{fig:5}. This smart strategy allows, with the manipulation of a single parameter, to realise both the compression and dilation of the energy spectrum, and the switching of the qubit-baths interactions. 

\begin{figure*}
		\begin{center}
		\includegraphics[width=\linewidth]{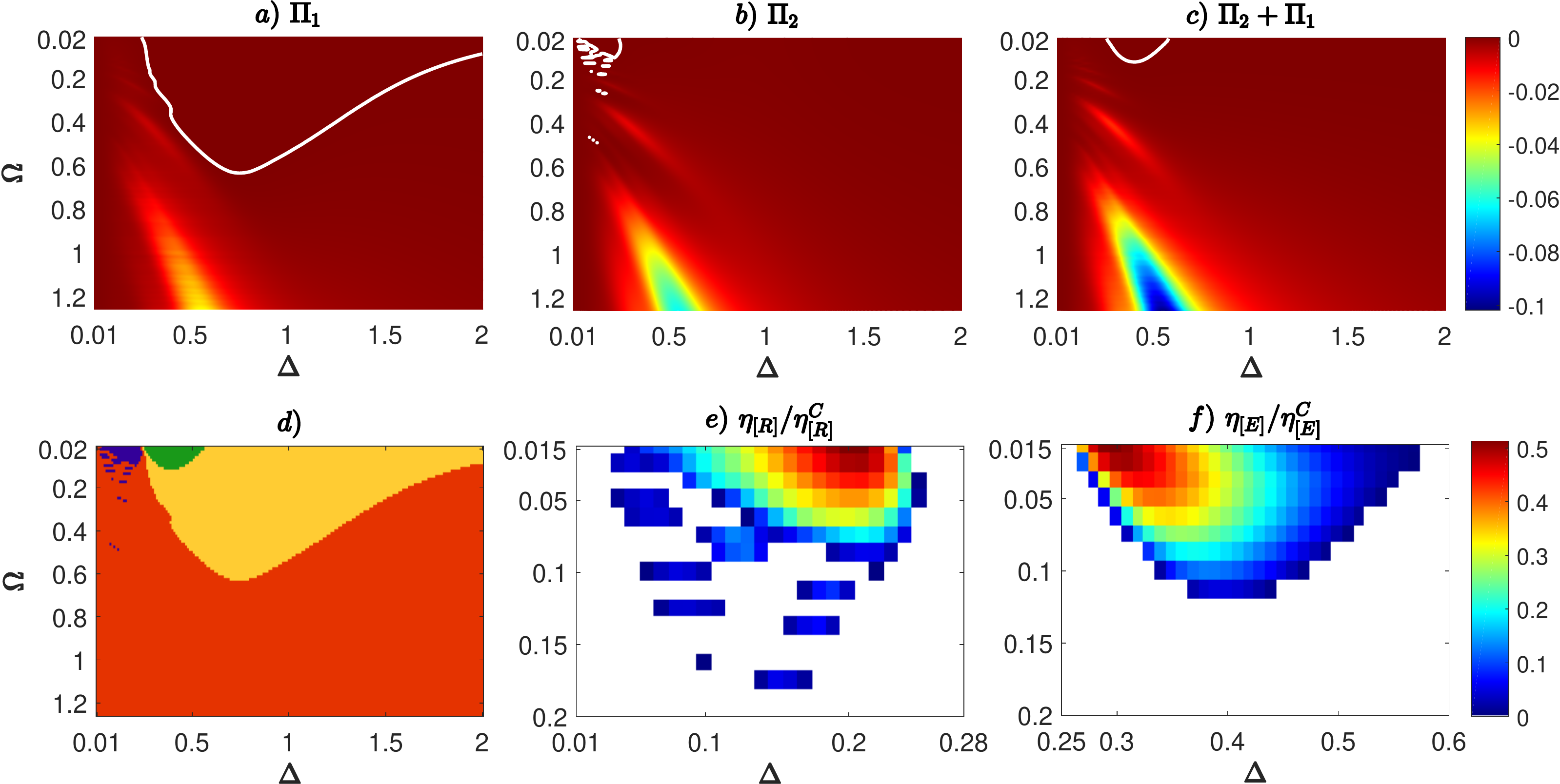}
		\caption{Thermodynamics of the quantum Otto engine of Karimi and Pekola \cite{Karimi16PRB94} as function of the parameters $\Omega,\Delta$. Panel a): Power withdrawn from resistor 1.  Panel b): Power withdrawn from resistor 2. Panel c): Power output. White curves denote the zero-level contours. Panel d): Operation regions: Blue=[R]; Red=[H]; Green = [E]; Yellow = [A], in accordance with the convention set in Fig. \ref{fig:1}a). Panel e): Rescaled refrigeration efficiency $\eta_{[R]}/\eta^C_{[R]}$. Panel f):  Rescaled heat engine efficiency $\eta_{[E]}/\eta^C_{[E]}$. The temperature ratio is $T_1/T_2=\beta_2/\beta_1=2$, while $\beta_1 E_0 =10/3 $. The Carnot point is accordingly at $\Omega=0, \Delta = 1/\sqrt{12}\simeq0.29$, and the Carnot efficiencies are $\eta_{[E]}^{C} = 1/2$ and $\eta_{[R]}^{C} =1 $. Following  Ref. \cite{Karimi16PRB94} the quality factor of both LRC circuits is set to the value $\mathcal{Q}_1=\mathcal{Q}_2=30$.}
		\label{fig:6}
		\end{center}
\end{figure*}
\begin{figure*}
		\begin{center}
		\includegraphics[width=\linewidth]{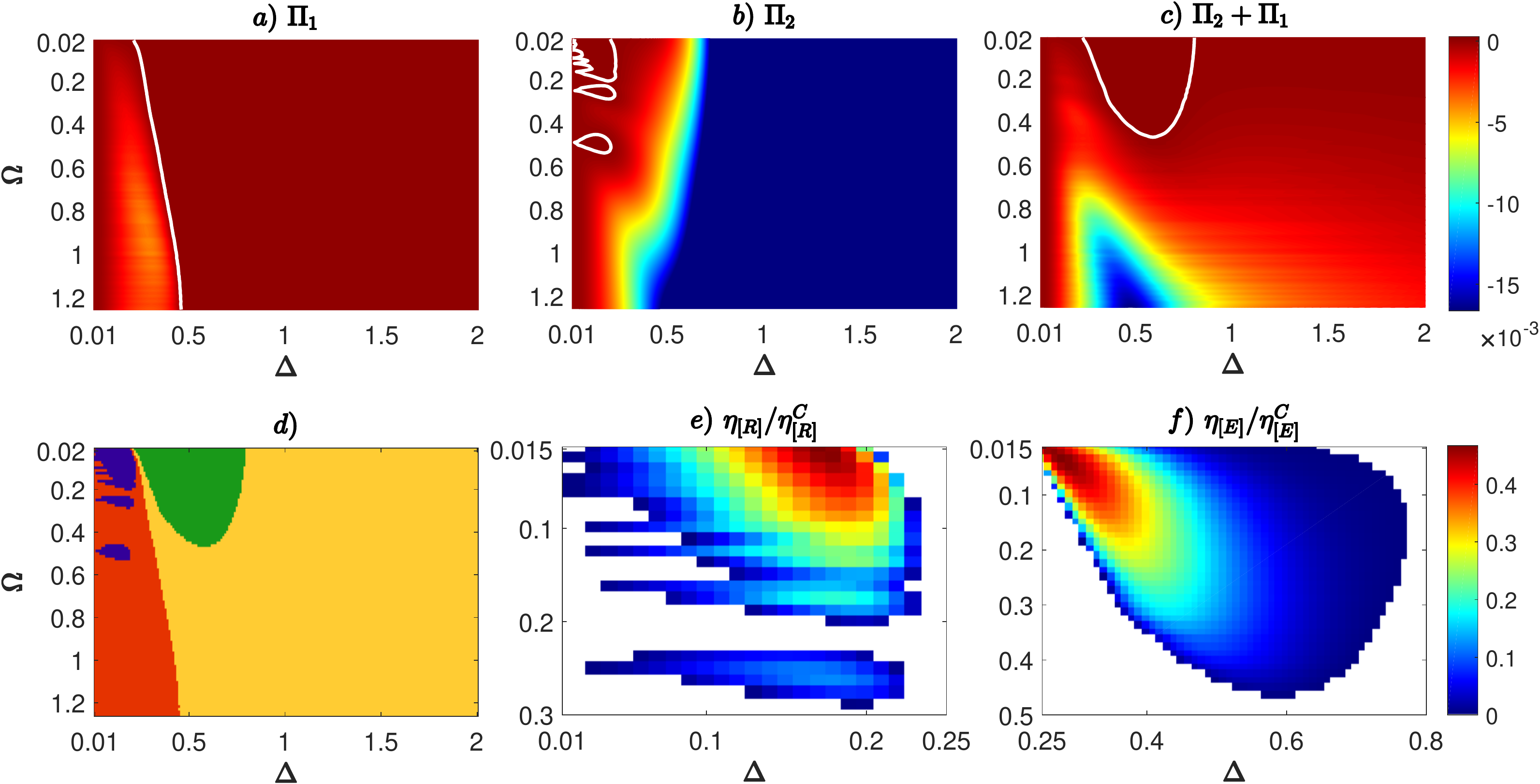}
		\caption{Same as Fig. \ref{fig:6} but for $\beta_1 E_0 =1/3$}
		\label{fig:7}
		\end{center}
\end{figure*}
The qubit Hamiltonian reads:
\begin{align}
H = -E_0 \Delta \sigma_x -E_0 q(t) \sigma_z\, ,
\end{align}
with 
\begin{align}
q(t) = \frac{1+\cos(2\pi f t)}{4}.
\end{align}
Accordingly the qubit level spacing periodically oscillates between the maximal value $\hbar \omega_1=2E_0 \sqrt{\Delta^2+1/4}$ and the minimal value $\hbar \omega_2=2E_0 \Delta$, where it is in resonance respectively with the hot and cold bath. Recall that $\Delta$ is a dimensionless quantity, while $E_0$ is an energy.

At variance with the idealised situation described in Sec.\ref{sec:LZSM}, in this case strokes 1 and 3 are not perfectly separated from 
the thermalisation strokes 2 and 4. Accordingly, they cannot be described by means of unitary evolution. Note also that the qubit remains in contact with the baths for a finite time which might not result in its full thermalisation. Following Karimi and Pekola \cite{Karimi16PRB94}, we describe the  dynamics of the open qubit, encompassing both the interaction with the resistors and the time-dependent driving, by means of the standard quantum master equation reading, in the instantaneous qubit energy eigenbasis:
\begin{align}
\dot\rho_{gg}(t) &= -\frac{\Delta}{q^2(t)+\Delta^2}\dot q(t) \Re [\rho_{ge}(t)e^{i\phi(t)}]-\Gamma_\Sigma \rho_{gg}(t)+\Gamma_\downarrow \\
\dot\rho_{ge}(t) &= \frac{\Delta}{q^2(t)+\Delta^2}\dot q(t) (\rho_{gg}(t)-1/2)e^{-i\phi(t)} - \Gamma_\Sigma \rho_{ge}(t)/2
\end{align}
where $\phi(t)=\int_0^t\omega(t')dt'$ is a dynamical phase determined by the instantaneous level spacing $\hbar \omega(t)=2 E_0\sqrt{q^2(t)+\Delta^2}$; $\Gamma_\downarrow$ denotes jump down transition rate of the qubit caused by the interaction with the two baths, i.e.,
$\Gamma_\downarrow= \Gamma_{\downarrow,1}+\Gamma_{\downarrow,2}$, with $\Gamma_{\downarrow,i}$ the jump down transition rate caused by bath $i$; $\Gamma_\Sigma=\sum_{j=1}^2 ( \Gamma_{\downarrow,j}+\Gamma_{\uparrow,j})$ being the sum of all transition rates caused by all the baths. 

The transition rates can be calculated by means of the Fermi golden rule leading to the expression \cite{Karimi16PRB94}:
\begin{align}
\Gamma_{\downarrow(\uparrow),j}=\frac{E_0^2M_j^2}{\hbar^2\Phi_0^2} \frac{\Delta^2}{q^2+\Delta^2}S_{I,j}(\pm \omega)
\label{eq:master}
\end{align}
where $S_{I,j}(\omega)=|R_j[1+\mathcal{Q}_j^2(\omega/\omega_{LC,j}-\omega_{LC,j}/\omega)]|^{-2}S_{V,j}(\omega)$ is the unsymmetrized noise spectrum of LRC circuit $j$ expressed in terms of its resonance frequency $\omega_{LC,j}=1/\sqrt{L_jC_j}$, quality factor $\mathcal{Q}_j=\sqrt{Lj/C_j}/R_j$ and voltage noise across the resistor, $S_{V,j}(\omega)=2R_j\hbar \omega/(1-e^{-\beta_j\hbar\omega}) $.
Here $M_j$ is the mutual inductance beteween the qubit and the $j$-th RLC circuit, and $\Phi_0=h/2e$ is the flux quantum ($e$ denotes the electron charge and $h$ is Planck's constant).

The power extracted from resistor $j$ by the qubit reads \cite{Karimi16PRB94} 
\begin{align}
\mathcal{P}_i(t)=- E(t) (\rho_{ee} \Gamma_{\downarrow,j}-\rho_{gg} \Gamma_{\uparrow,j}).
\end{align}
We present results for the time averaged powers, at steady state, in dimensionless form,
\begin{align}
\Pi_i= \frac{\hbar}{E_0^2 T}\int_t^{t+T} \Pi_i(s)ds \, ,
\end{align}
as a function of $\Delta$ and the dimensionless drive frequency 
$
\Omega= {2\pi \hbar f}/{E_0}
$
where $T=1/f$ is the driving period.

In Fig. \ref{fig:6} we plot $\Pi_i, i=1,2$, the according power output $\Pi_W=\Pi_1+\Pi_2$, the operation regions and the $[E]$ and $[R]$ efficiencies (rescaled by the according Carnot efficiencies), as functions of $\Delta$ and $\Omega$, for fixed ratio $\beta_1/\beta_2$. Same in Fig. \ref{fig:7} but for higher temperatures. 

Note how some oscillations, signaling the presence of quantum coherences, are present in the low temperature power plots, Fig. (\ref{fig:6}), while they are washed away in the higher temperature plots, Fig. (\ref{fig:7}). This effect is analogous to that observed in the ideal case, Figs. (\ref{fig:3},\ref{fig:4}). The oscillations are as well reflected in the plots of the regions of operation resulting in a breakdown of the regions connectedness, see Panel d) of Figs. (\ref{fig:6},\ref{fig:7}).

Inspection of Panels e) and f) of both Fig. \ref{fig:6} and \ref{fig:7} shows, as expected, that the efficiencies grow as the Carnot point is approached, note however that both in the [R] and [E] regions, the maximal efficiency is about half the Carnot efficiency.

Note that, in the large $\Delta$ limit, only the [H] operation occurs. While the physics behind this phenomenon can be physically understood based on the previous analysis based on the idealised LZSM dynamics, its emergence can also be seen analytically by taking the large $\Delta$ limit of the master equation (\ref{eq:master}), see Appendix \ref{app:C}.

\section{Remarks and Conclusions}
\label{sec:conclusions}
We have illustrated the emergence of the 4 operation modes allowed by the laws of thermodynamics, in a single qubit quantum Otto engine operating in finite time. We have begun with a general treatment of the idealised case where thermalisation and generic thermally-insulated strokes  are well separated. The geometrical properties of the various operation regions in the ($\omega_2/\omega_1,P$) space is all encoded into the bath temperatures, where $P$ is the transition probability among the qubit levels during the thermally-insulated strokes, and $\omega_{1,2}$ are the qubit level spacings during the thermalisation strokes. Simple analytical expressions have been obtained for the boundaries of the regions. We have then specialised to the case of Landau-Zener-St\"uckelberg-Majorana dynamics. In this case coherent oscillations break the connectedness of the operation mode regions in the parameter space $(\Delta,v)$. 

We then investigated the realistic engine proposed by Karimi and Pekola \cite{Karimi16PRB94}. We have, accordingly, provided a fully fledged characterisation of the operation of that device in its parameter space $(\Delta,\Omega)$, which constitute a solid basis for its design and practical realisation. The study of the idealised Landau-Zener-St\"uckelberg-Majorana (LZSM- in short) dynamics provides a good guide to understand its physics through an exactly solvable simplified model. In the realistic case of Karimi and Pekola, coherent effects are less evident than in the idealised LZSM case, which is probably a consequence of a smoother driving (rather than of the continuous interaction with the baths  \cite{Karimi16PRB94}).

 We remark that the connectedness of the operation mode regions is important because it is associated with the robustness of the engine operation against experimental noise on the control parameters. In this regard our work corroborates the finding of Karimi and Pekola \cite{Karimi16PRB94} that operation at too low temperature values might be hindered by coherences. We also noted that despite with increasing energy gap $E_0 \Delta$ the quantum adiabatic approximation gets better and better, the engine tends to become a mere heater as $\Delta \to \infty$ limit, and have explained the origin of this phenomenon.

\appendix
\section{Operation modes allowed by Clausius inequality for the two-baths case}
\label{app:A}
Eqs. (\ref{eq:clausius}, \ref{eq:NRGconservation}), combined with the convention 
\begin{align}
0 < \beta_1 < \beta_2
\label{eq:beta-convention}
\end{align}
are incompatible with four sign combinations for $Q_1,Q_2,W$.

The case $W > 0$, $Q_1< 0$, $Q_2 < 0$ is not allowed because if $Q_1> 0$, $Q_2> 0$, then, by Eqs. (\ref{eq:NRGconservation},\ref{eq:beta-convention}), it must be $W>0$.

The case $W < 0$, $Q_1>0$, $Q_2 > 0$ is not allowed because if  $Q_1>0$, $Q_2 > 0$, then, by Eqs. (\ref{eq:NRGconservation},\ref{eq:beta-convention}), it must be  $W>0$.

The above two cases are not consistent with energy conservation, Eq. (\ref{eq:NRGconservation}).

Regardless of the sign of $Q_1$, the case $W>0,Q_2>0$ is incompatible with the Clausius inequality  (\ref{eq:ClausiusFor2}).
In fact assuming $W>0,Q_2>0$ it is, since $\beta_1>0$, $0<\beta_1 Q_1 +\beta_1 Q_2< \beta_1 Q1+ \beta_2 Q_2$, because of Eq. (\ref{eq:beta-convention}), which is in disagreement with Eq. (\ref{eq:ClausiusFor2}), hence both the case $W > 0$, $Q_1>0$, $Q_2 > 0$, and the case  $W > 0$, $Q_1<0$, $Q_2 > 0$ are not allowed. This reflects the impossibility of having a machine that is at the same time a heat engine (i.e., a work provider $W>0$) and a refrigerator ($Q_2>0$).

No incompatibility exists for the remaining four cases listed in Eqs. (\ref{eq:R},\ref{eq:E},\ref{eq:A},\ref{eq:H}).

\section{Clausius inequality for an ideal one-qubit quantum Otto engine}
\label{app:B}
Setting $\hbar \omega_1 \beta_1/2=x$, $\hbar \omega_2 \beta_2/2=y$, the  Clausius sum $\Sigma=  \beta_1 Q_1+\beta_2 Q_2$, reads, with $Q_{1,2}$ from Eqs. (\ref{eq:Q1},\ref{eq:Q2}):
\begin{align}
\Sigma =-  y \tanh y - x \tanh x + (2P-1) (y \tanh x+x \tanh y)
\end{align}
Since $x,y\geq 0$, it is $\tanh x\geq 0, \tanh y \geq 0$, hence $y \tanh x+x \tanh y \geq 0$. Since $ 2P-1 \leq 1$, it is $(2P-1)(y \tanh x+x \tanh y) \leq (y \tanh x+x \tanh y)$. Therefore:
\begin{align}
\Sigma &\leq  - y\tanh y - x \tanh x + y \tanh x+x \tanh y \nonumber \\
&= (y-x)(\tanh  x- \tanh y) \leq 0
\end{align}
where the last inequality follows from the fact that the hyperbolic tangent is a monotonously increasing function. 
Note that $\Sigma$ vanishes, that is it gets its largest possible value, at the Carnot point $x=y, P=1$ (namely $\omega_1 \beta_1=\omega_2 \beta_2, P=1$).

\section{Master equation in the large $\Delta$ limit}
\label{app:C}
In this appendix we solve the master equation describing the open qubit evolution in the limit $\Delta\gg 1$. This will illustrate the mechanism underling the fact that in the large $\Delta$ region of parameters only the heater [H] operation is possible even if the adiabatic approximation is better achieved when $\Delta$ gets larger. The transition rates $\Gamma_{\downarrow(\uparrow),j}$ read with the $\Delta$ dependence made explicit
\begin{align}
\Gamma_{\downarrow,1} = A_1 \frac{\Delta^2}{q^2+\Delta^2}
&\left|1+\mathcal{Q}_1\left(\sqrt{\frac{q^2+\Delta^2}{{\bar q}^2+\Delta^2}}-\sqrt{\frac{{\bar q}^2+\Delta^2}{q^2+\Delta^2}}\right)\right|^{-2} \nonumber \\
&\times\frac{\sqrt{q^2+\Delta^2}}{1-e^{-2E_0 \beta_1\sqrt{q^2+\Delta^2}}}\\
\Gamma_{\downarrow,2} = A_2 \frac{\Delta^2}{q^2+\Delta^2}
&\left|1+\mathcal{Q}_2\left(\frac{\sqrt{q^2+\Delta^2}}{\Delta}-{\frac{\Delta}{\sqrt{q^2+\Delta^2}}}\right)\right|^{-2} \nonumber \\
& \times\frac{\sqrt{q^2+\Delta^2}}{1-e^{-2\beta_2E_0\sqrt{q^2+\Delta^2}}}\\
\Gamma_{\uparrow,1} = -A_1 \frac{\Delta^2}{q^2+\Delta^2}
&\left|1+\mathcal{Q}_1\left(\sqrt{\frac{q^2+\Delta^2}{{\bar q}^2+\Delta^2}}-\sqrt{\frac{{\bar q}^2+\Delta^2}{q^2+\Delta^2}}\right)\right|^{-2} \nonumber \\
&\times \frac{\sqrt{q^2+\Delta^2}}{1-e^{2E_0 \beta_1\sqrt{q^2+\Delta^2}}}\\
\Gamma_{\uparrow,2} =- A_2 \frac{\Delta^2}{q^2+\Delta^2}
&\left|1+\mathcal{Q}_2\left(\frac{\sqrt{q^2+\Delta^2}}{\Delta}-{\frac{\Delta}{\sqrt{q^2+\Delta^2}}}\right)\right|^{-2} \nonumber \\
&\times\frac{\sqrt{q^2+\Delta^2}}{1-e^{2\beta_2E_0\sqrt{q^2+\Delta^2}}}
\end{align}
where $q$ is a shorthand for $q(t)$, $\bar q= \max q(t)$ and
$A_1$ and $A_2$ are factors with the dimension of frequency,  that contain information about the two resonators and the qubit energy scale $E_0$. Performing a Taylor expansion up to the leading order in $1/\Delta$ we obtain
\begin{align} 
\Gamma_{\downarrow,1}&\simeq A_1\Delta\left(1-\frac{1}{2}\left(\frac{q}{\Delta}\right)^2\right)\times \nonumber \\
&\left(1+2\mathcal{Q}_1\frac{({\bar q}^2-q^2)}{\Delta^2}\right)\left(1+e^{-2E_0 \beta_1\Delta}\right) 
\simeq A_1\Delta \\
\Gamma_{\downarrow,2} & \simeq A_2\Delta\left(1-\frac{1}{2}\left(\frac{q}{\Delta}\right)^2\right)\times \nonumber \\
&\left(1-2\mathcal{Q}_2\left(\frac{q}{\Delta}\right)^2\right)\left(1+e^{-2\beta_2E_0\Delta}\right) 
\simeq A_2\Delta \\
\Gamma_{\uparrow,1}&\simeq -A_1\Delta\left(1-\frac{1}{2}\left(\frac{q}{\Delta}\right)^2\right)
\times \nonumber \\
&
\left(1+2\mathcal{Q}_1\frac{({\bar q}^2-q^2)}{\Delta^2}\right)e^{-2E_0 \beta_1\Delta}  \simeq -A_1\Delta e^{-2E_0 \beta_1\Delta}\\
\Gamma_{\uparrow,2}&\simeq -A_2\Delta\left(1-\frac{1}{2}\left(\frac{q}{\Delta}\right)^2\right)
\times \nonumber \\
&
\left(1-2\mathcal{Q}_2\left(\frac{q}{\Delta}\right)^2\right)e^{-2\beta_2E_0\Delta}\simeq -A_2\Delta e^{-2\beta_2E_0\Delta}
\end{align}
The master equations for the time evolution of $\rho_{gg}$ and $\rho_{ge}$ then reads
\begin{align}
\dot\rho_{gg}(t) &= -\frac{\Delta}{q^2(t)+\Delta^2}\dot q(t) \Re [\rho_{ge}(t)e^{i\phi(t)}]-\Gamma_\Sigma \rho_{gg}(t)+\Gamma_\downarrow  \nonumber \\
&\simeq (A_1+A_2)\Delta(1-\rho_{gg})\\
\dot\rho_{ge}(t) &= \frac{\Delta}{q^2(t)+\Delta^2}\dot q(t) (\rho_{gg}(t)-1/2)e^{-i\phi(t)}- \Gamma_\Sigma \rho_{ge}(t)/2 \nonumber \\
&\simeq-(A_1+A_2)\Delta\rho_{ge}/2
\end{align}
where $\Gamma_\Sigma =\Gamma_{\downarrow,1}+\Gamma_{\downarrow,2}+\Gamma_{\uparrow,1}+\Gamma_{\uparrow,2}\simeq(A_1+A_2)\Delta$ and $\Gamma_{\downarrow}=\Gamma_{\downarrow,1}+\Gamma_{\downarrow,2}\simeq (A_1+A_2)\Delta$ and the fist terms on the right hand side of both equations have been neglected because they are of order $1/\Delta$ while all the other terms are of order $\Delta$. In this limit the two equation are no more coupled and they can be easily solved:
\begin{align}
\rho_{gg}(t) &\simeq 1  - (1-\rho_{gg}(0))e^{-\Delta(A_1+A_2)t}
\label{rhoggt}\\
\rho_{ge}(t) &\simeq \rho_{ge}(0)e^{-\Delta(A_1+A_2)t/2}
\label{rhoget}
\end{align}
In our case the initial state of the qubit at $t = 0$ is a thermal state at reverse temperature $\beta_1$ and when $\Delta\gg\ 1$ it becomes
\begin{align}
\rho_{gg}(0)&=\frac{e^{E_0 \beta_1\sqrt{{\bar q}^2+\Delta^2}}}{\cosh(E_0 \beta_1\sqrt{{\bar q}^2+\Delta^2})}\simeq\frac{1}{1+e^{-2E_0 \beta_1\Delta}}\nonumber \\
&\simeq 1-e^{-2E_0 \beta_1\Delta}\label{rhogg0}\\
\rho_{ge}(0) &\simeq 0
\label{rhoge0}
\end{align}
Substituting (\ref{rhogg0}) and (\ref{rhoge0}) into equation (\ref{rhoggt}) and (\ref{rhoget}) respectivly we obtain
\begin{align}
&\rho_{gg}(t) \simeq 1-e^{-\Delta(E_0 \beta_1+(A_1+A_2)t)}\\
&\rho_{ge}(t) \simeq 0
\end{align}
We note that for $\Delta\rightarrow\infty$ it is $\rho_{gg}\rightarrow 1$. Accordingly the qubit tends to stay in its ground state for all $t$'s and this effect is more evident for larger $\beta_1$ (smaller temperatures).
The expression for instantaneous dimensionless power $\Pi_j$ to resistor $j$ reads
\begin{align}
\Pi_j(t)&= 2 (\hbar/E_0) \sqrt{q^2+\Delta^2}(\Gamma_{\downarrow,j}-\rho_{gg}(\Gamma_{\downarrow,j}+\Gamma_{\uparrow,j}))\nonumber \\
&\simeq 2 (\hbar/E_0)\Delta\left(A_j\Delta -\left( 1-e^{-\Delta(E_0 \beta_1+(A_1+A_2)t)}\right)A_j\Delta\right) \nonumber\\
&\simeq 2 (\hbar/E_0) A_j\Delta^2e^{-\Delta(E_0 \beta_1+(A_1+A_2)t)}\geq 0
\end{align}
Hence the  dimensionless average value of power over a period $T$ of the driving become
\begin{align}
\Pi_j &\simeq\frac{2\hbar A_j\Delta^2}{E_0T} \int_{0}^{T}e^{-\Delta(E_0 \beta_1+(A_1+A_2)s)}ds \nonumber\\ 
&=\frac{2\hbar A_j}{E_0^2T}\frac{(1-e^{-E_0 \beta_1\Delta(A_1+A_2)T})}{ \beta_1(A_1+A_2)}\Delta e^{-E_0 \beta_1\Delta} \nonumber\\
&\simeq \frac{2\hbar}{E_0^2T}\frac{A_j}{A_1+A_2}\frac{\Delta}{ \beta_1}e^{-E_0 \beta_1\Delta}\geq 0
\end{align}
Accordingly in the large $\Delta$ limit,  both the powers to resistors are positive and consequently the only possible regime is the heater [H].

\end{document}